\documentstyle[espart12]{article}
\def\t2{\tilde t}
\def\u2{\tilde u}
\def\s2{\tilde s}
\def\J{J/\psi}
\def\MJ{M_{\psi}}
\def\mc{m_c}
\def\cg{c(\gamma)g\to J/\psi c}
\def\gg{g(\gamma)g\to J/\psi g}

\def\als{\alpha_s}
\def\sq{\sqrt{s_{\gamma p}}}
\def\pt{p_{\bot}}
\begin{document}
\begin{center}
 {\bf CHARM CONTENT OF A PHOTON AND $\J$
   PHOTOPRODUCTION AT HIGH ENERGIES}

 V.A.Saleev

Samara State University, Samara, 443011, Russia

\end{center}
\begin{abstract}
A study of the $\J$-meson production at large transverse momentum
in high energy $\gamma p$ collisions via the charm quark excitation in
a photon is presented. Based on perturbative QCD and the
nonrelativistic quark model, our calculation demonstrates that the
charm content of a photon may be very important for the $\J$
photoproduction at
the large transverse momentum. It is shown that at HERA energies
$\J$ production via the subprocess $\cg$ dominates over the resolved
photon contribution via the subprocess $\gg$ at $\pt>3$ GeV/c
and over the direct photon-gluon fusion contribution  at
$\pt>10$ GeV/c.
\end{abstract}

\section{Introduction} The  study  of  a  different  mechanisms of the
inclusive  $\J$  photoproduction  on  protons  is  of   a   particular
importance  because  the  process  of $\J$ production via photon-gluon
fusion plays a crucial role in  the  measurement  of  gluon  structure
function in  a  proton  \cite{1,2,3,4}.  Previous calculations of $\J$
photoproduction at   large   $\pt$   have  included  direct  charmonium
production via photon-gluon fusion  using  the  colour  singlet  model
\cite{1,2}, diffractive  inelastic $\J$ production \cite{5},  resolved
photon contributions via subprocesses  $gg\to\J g$,
$gg\to\chi_{0,2}\to\J\gamma$ \cite{4} and $\J$ production from
$b$ quark  decays  \cite{6}.  Here  we  examine  the  diffractive-like
contribution of  the  charm quark excitation in a photon to
the $\J$ photoproduction at large $\pt$ via partonic subprocess $\cg$,
where charm  quarks  in  the  initial  state  are not pure "intrinsic"
\cite{7} but are generated by QCD evolution of  the  photon  structure
functions (PSF)  \cite{8}.  We  suppose  that  at  low $Q^2$ the charm
content of the photon is virtually nil,  but at $Q^2$ of order $m_c^2$
one has  sufficient  resolution to find charm quarks in photon.  Note,
that in the processes of the $\J$ photoproduction at large  $\pt$  the
relevant QCD scale $Q^2\sim \MJ^2+\pt^2>>m_c^2$.  Our calculation
is partly  the  same  as  the  approach  used  in   Ref.\cite{9}   for
description of  the  charm quark hadroproduction and based on the fact
that both the central and diffractive components of  charm  production
can be understood in the context of perturbative QCD.  Recently we have
presented the results of the calculations  for  contributions  of  the
intrinsic charm  in  the  proton to the $\J$ photoproduction \cite{10}
and hadroproduction \cite{11}.

The study of a photon structure function \cite{8} in the
resolved-photon interaction at HERA energy and beyond is also very
interested \cite{12}. Usually it is supposed that at high energy in
the resolved-photon interaction the gluon content of a photon is
dominant. In this paper we show that in the $\J$ photoproduction
at lager $\pt$ via
resolved-photon interaction the charm content of a photon is more
important and it is may be studied experimentally using $\J$ plus
open charm associated photoproduction.

\section{Model}
The process of the $\J$ photoproduction on a proton in the resolved-photon
interaction is schematically presented in Fig. 1.
We start from the lowest order in $\als$ amplitudes which correspond the set
of Feynman diagrams in Fig. 2. As usual in potential model, the quarkonium
is represented as nonrelativistic quark-antiquark bound system in singlet
colour state with specified mass $\MJ=2\mc$ ( $\mc$ is charm quark mass) and
spin-parity
$J^p=1^-$. The amplitudes for subprocess $\cg$ can be expressed in the
following form:
\begin{equation}
 M_1=g^3T^cT^aT^a\bar U(q')\hat\varepsilon_{g}\frac{\hat q'-\hat k+\mc}
 {(k-q')^2-\mc^2}\gamma^{\mu}\frac{\hat P}{(p-q)^2}
  \gamma_{\mu}U(q),
\end{equation}
\begin{equation}
M_2=g^3T^aT^aT^c \bar U(q')\gamma^{\mu}\frac{\hat P}{(p+q')^2}
\gamma_{\mu}
\frac{\hat k+\hat q+\mc}{(k+q)^2-\mc^2}\hat\varepsilon_{g}U(q),
\end{equation}
\begin{equation}
M_3=g^3T^aT^cT^a\bar U(q')\gamma^{\mu}\frac{\hat P}{(q'+p)^2}\hat
\varepsilon_g
\frac{\hat p-\hat k+\mc}{(p-k)^2-\mc^2}\gamma_{\mu}U(q),
\end{equation}
\begin{equation}
M_4=g^3T^aT^cT^a\bar U(q')\gamma^{\mu}
\frac{\hat k-\hat p+\mc} {(p-k)^2-\mc^2}\varepsilon_{g}
\frac{\hat P}{(p-q)^2}\gamma_{\mu}U(q),
\end{equation}
\begin{equation}
 M_5=-ig^3f^{bac}T^bT^a\bar U(q')\gamma^{\rho}\frac{\hat P}{(p+q')^2}
 \frac{\gamma^{\sigma}}{(q-p)^2}U(q)C_{\mu\sigma\rho}\varepsilon^
  {\mu}_g,
\end{equation}
where
$$C_{\mu\sigma\rho}=(q+q')_{\mu}g_{\sigma\rho}-(p+k+q')_{\sigma}
  g_{\mu\rho}+(k-q+p)_{\rho}g_{\mu\sigma}.$$
The spin and colour properties of the $c\bar c$-bound system is described
using projection operator \cite{1,2}:
\begin{equation}
\hat P=\frac{F_c\Psi(0)}{2\sqrt {\MJ}}\hat\varepsilon_J(\hat p_{\psi}+\MJ),
\end{equation}
where  $F_c=\delta^{kr}/\sqrt{3}$,  k and  r  are  colour
indexes of charm quarks,
$T^a=\lambda^a/2$. The value of $\J$ wave function at the origin $\Psi(0)$
can be extracted  in
the lowest  order  of perturbative QCD from the leptonic decay
width of the $J/\psi$:
\begin{equation}
  \Gamma_{ee}=4\pi e_q^2\alpha^2\frac{|\Psi(0)|^2}{\mc^2}.
\end{equation}
We shall put in our calculation $\als=0.3$, $\mc=1.55$ GeV,
$\Gamma_{ee}=5.4$ KeV \cite{9}.

After average and sum over spins and  colours  of  initial  and  final
particles, we obtain the expression for square of matrix element:
  \begin{equation}
  \overline{|M|^2}=B_{gc}\sum_{i\le j=1}^{5}C_{ij}K_{ij}(\hat s,\hat t,\hat u),
   \end{equation}
 where
  $$B_{gc}=\frac{\pi^2\als^3\Gamma_{ee}}
    {16\alpha^2\mc},$$
$\hat s=(k+q)^2,\qquad \hat t=(k-q')^2,\qquad \hat u=(q-q')^2$
are usual Mandelstam variables
and $\hat s+\hat t+\hat u=6m^2$. The explicit analytical formula for
functions $K_{ij}$ are given in Appendix.

The appropriate colour factors are:
\begin{eqnarray}
&&C_{11}=C_{22}=C_{12}=C_{21}=64/9,\nonumber\\
&&C_{33}=C_{44}=C_{34}=C_{43}=1/9,\nonumber\\
&&C_{13}=C_{31}=C_{14}=C_{41}=C_{23}=C_{32}=C_{24}=C_{42}=-8/9,
  \nonumber\\
&&C_{15}=C_{51}=C_{25}=C_{52}=8,\nonumber\\
&&C_{35}=C_{53}=C_{45}=C_{54}=-1, \qquad  C_{55}=-9
\end{eqnarray}

The differential cross section for subprocess $\cg$
can be written as follows:
\begin{equation}
 \frac{d\hat\sigma}{d\hat t}=\frac{\overline{|M|^2}}{16\pi(\hat s-\mc^2)^2}
  \end{equation}

In the conventional parton model the measurable cross-section
is obtained by folding the hard parton level cross-section  with
the respective parton densities:
\begin{eqnarray}
\frac{d\sigma}{d^2\pt dy}&=&\int dx_1\int dx_2
C_{\gamma}(x_1,Q^2)G_p(x_2,Q^2) \nonumber\\
&& \frac{d\hat\sigma}{d\hat t}
  (cg\to\J c)\frac{x_1x_2 s}{\pi}\delta(\hat s+\hat t+\hat u-\MJ^2
-2\mc^2).
\end{eqnarray}
Here: $\hat s=x_1x_2s+\mc^2,\quad \hat t=\MJ^2+\mc^2-x_1\sqrt{s}
M_{\bot}\exp(-y^{\star}),\quad \hat
u=\MJ^2-x_2\sqrt{s}M_{\bot}\exp(y^{\star}),\quad
M_{\bot}=\sqrt{\MJ^2+\pt^2},$
where $y$ is the $\J$ rapidity in c.m.s., $\pt$ is the $\J$ transverse
momentum, $G_p(x_2,Q^2)$ is the gluon distribution function in a proton,
at the scale $Q^2=M_{\bot}^2$, $C_{\gamma}(x_1,Q^2)$ is the charm quark
distribution function in a photon, $s$ is the square of a total energy of
colliding particles in the $\gamma p$ center of mass reference frame.
We use in calculations LO GRV \cite{14} parameterization for
$G_p(x_2,Q^2)$ structure function. The input charm quark distribution
$C_{\gamma}(x_1,Q^2)$ demands the more careful consideration.

In all QCD motivated parton distributions of the photon \cite{15}
heavy quarks are considered as an intrinsic massless parton whose
distribution is generated according to the Altarelli--Parisi
equations. In spite of the fact that in the process of the $\J$
production at large $\pt$ the relevant scale $Q^2>>\mc^2$ and a
resummation procedure of the perturbation series $(\alpha_s\log
(Q^2/\mc^2)^n$ is valid, we don't use these parameterizations
taking into account that our consideration based on the full massive
subprocesses. As it was shown in Ref. \cite{16}, the concept of the
massless intrinsic heavy quark distributions yield a overestimated
result for heavy quarks production at high energies. That is why more
grounded to use "hard" phenomenological parameterization based on the
vector-meson-dominance (VMD) model.
In the approach of VMD model the c-quark PSF is presented via the
charm quark
distribution function of the $J/\psi$ meson:
\begin{equation}
 C_{\gamma}(x)=k\frac{4\pi\alpha}{f^2_{\psi}}C_{\psi}(x),
\end{equation}
with $1\leq k\leq 2$. The precise value of $k$ clearly has to be extracted
from experiment. Similar to Ref. \cite{17}, where the photoproduction
of charm hadrons in VMD model has been discussed, we use for
distribution charm quark in $\J$-meson the simple scaling
parameterization
\begin{equation}
 C_{\psi}(x)=49.8 x^{1.2}(1-x)^{2.45}.
\end{equation}
We want to note that this parameterization takes into account heavy quark
mass effects \cite{17}, which are principal for our consideration.

Let us next consider the dynamical cutoff for the charm excitation
processes. For the typical charm excitation diagram (Fig.2), the charm
quark must receive sufficient transfer momentum, which is necessary
to excite $c\bar c$ pair. This implies a minimum dynamical resolution
$|\hat t|_{min}$ for the momentum transfer $\hat t$ of the $\cg$
subprocesses. We choose $|\hat t|_{min}=\mc^2$, although the
specification of the scale is uncertain by factors \cite{9}. The
dependence of the our results from the choice of $|\hat t|_{min}$ will
be discussed later.

\section{Results and discussion}
Because of the relevant QCD scale $Q^2$ is order of $M_{\bot}$, we
consider at first $\pt$ distribution of the $\J$-mesons which are
generated in $\gamma p$ collisions via charm excitation in the photon.
Fig.~3 shows our predictions for $\J'$s $\pt$-spectra at the energy
$\sq=200$ GeV. The solid curve corresponds to the contribution of the
photon-gluon fusion mechanism. The short-dashed curves are
contributions of the charm excitation subprocesses at different choice
of dynamical cutoff: upper curve corresponds to $|\hat
t|_{min}=\mc^2$, lower curve - $|\hat t|_{min}=4\mc^2$. The
long-dashed curve is the contribution of the resolved photon
interaction via subprocess $\gg$. We can see that the result of
calculation for charm excitation mechanism is independent from $|\hat
t|_{min}$ at $\pt>5$ GeV/c. But in low $\pt$ region one has strong
suppression of the $\pt$-spectrum versus $|\hat t|_{min}$. So, our
predictions, based on the charm excitation mechanism, are realistic for
large $\pt$ where $|\hat t|_{min}$ cutoff influence is absent.

Fig.~3 demonstrates the crossing of the solid and short-dashed curves at
$\pt=10$ GeV/c and at large $\pt>10$ GeV/c the charm excitation mechanism
dominates over the direct photon-gluon fusion. The contribution of the
resolved photon interaction via subprocess $\gg$ is bigger the
contribution of the $\cg$ subprocess only at $\pt<2$ GeV/c. The
rapidity distribution, which are presented in Fig.~4, also shows us
the sufficient role of the charm excitation process in the $\J$
photoproduction at large $\pt$. We see that the contribution of this
one in the $\J$'s rapidity spectrum at $\pt>7$ GeV/c is dominant in
the wide rapidity region $-0.7<y^{\star}<2.0$.

In the Fig.~5 the results of calculation for the total cross section
of the $\J$ photoproduction at $\pt>7$ GeV/c versus $\sq$ are presented.
The contribution of the subprocess $\cg$, which is independent of the
dynamical cutoff at large $\pt$, rapidly grows beginning with $\sq=40$
GeV and at $\sq=200$ GeV one has $\sigma(\gamma g\to\J g)/\sigma(\cg)
\approx 1.8$. At the energies $\sq>200$ GeV the behaviour of the total
cross sections both for the photon-gluon fusion as for the charm
excitation diagrams are the same and to be conditioned by the gluon
distribution function in the proton. The contribution of the resolved
photon interaction via subprocess $\gg$ in the total cross section at
large $\pt$ is two order of magnitude smaller the contributions of the
discussed above mechanisms.

Note, that we don't take into
consideration so-called $K$-factor which is needed as usual for
normalization of the leading order QCD predictions and experimental
data. In the present paper we accurately predict only relative
contributions of the different $\J$ production mechanism at large
$\pt$. The discussed here $\J$ photoproduction mechanism via the
charm quark excitation can be used also for prediction of the open
charm production rates in $\gamma p$ collisions at HERA energy and
beyond \cite{18}.

{\it\bf Acknowledgements.}

  Author thank  A.~Likhoded and A.~Martynenko for useful discussions.
  This  research was supported by the Russian
  Foundation of Basic Research (Grant 93-02-3545)
  and by State Committee on High Education of Russian Federation
  (Grant 94-6.7-2015).

{\large\bf Figure captions.}

\begin{enumerate}
\item The $\J$ photoproduction in resolved-photon interaction.
\item Diagrams used to describe the partonic subprocess $\cg$.
\item The $\pt$ distribution for $\J$ photoproduction at $\sq=200$ GeV
and all $y^{\star}$. The solid curve is the direct photon-gluon fusion
contribution. The long-dashed curve is the resolved photon
contribution via the $\gg$ subprocess. The short-dashed curves are the
contribution of the charm quark excitation in the photon, the upper
curve corresponds to $|\hat t|_{min}=\mc^2$ and the lower curve
corresponds to $|\hat t|_{min}=4\mc^2$.
\item The $y^{\star}$ distribution for the $\J$ photoproduction at
$\sq=200$ GeV and $\pt>7$ GeV/c. Notation as in Fig.~3.

\item The total cross section for the $\J$ photoproduction
 at $\pt>7$ GeV/c versus $\sq$. Notation as in Fig.~3.
 \end{enumerate}

\newpage
{\large\bf Appendix.}

$$\t2=\hat t/m^2,\quad \u2=\hat u/m^2,\quad \s2=\hat s/m^2$$
   \begin{eqnarray}
      K_{11}=-4(2\s2\t2-2\s2+\t2^2\u2-4\t2^2-8\t2\u2+14
      \t2+7\u2-106)/(\t2-1)^4
\end{eqnarray}
\begin{eqnarray}
      K_{12}&=&4(\s2^3-\s2^2\t2-6\s2^2-\s2\t2^2-2\s2\t2\u2+16
      \s2\t2-\s2\u2^2+8\s2\u2\nonumber\\
      && -28\s2+\t2^3-6\t2^2-\t2\u2
      ^2+8\t2\u2-28\t2+4\u2^2-126\u2+276)/\nonumber\\
      && ((\s2-1)^2(\t2-1)^2)
\end{eqnarray}
\begin{eqnarray}
      K_{13}&=&8(\s2^2\t2+\s2^2+2\s2\t2^2+\s2\t2\u2-22\s2\t2+7
      \s2\u2+16\s2+\nonumber\\
&&  2\t2^2\u2-24\t2^2-11\t2\u2+217
 \t2-2\u2^2+41\u2-511)/\nonumber\\
 && ((\s2-1)(\t2-1)^2(\u2-4))
\end{eqnarray}
\begin{eqnarray}
      K_{14}&=&-8(\s2^3+\s2^2\t2-18\s2^2+\s2\t2^2+\s2\t2\u2-16
      \s2\t2-\s2\u2^2-3\s2\u2+\nonumber\\
       &&166\s2-11\t2^2-5\t2\u2+
 115\t2+13\u2^2-65\u2-335)/\nonumber\\
 &&((\t2-1)^3(\u2-4))
\end{eqnarray}
\begin{eqnarray}
      K_{22}=-4(\s2^2\u2-4\s2^2+2\s2\t2-8\s2\u2+14\s2-2
      \t2+7\u2-106)/ (\s2^4-1)^4
\end{eqnarray}
\begin{eqnarray}
      K_{23}&=&-4(\s2^2\t2-11\s2^2+\s2\t2^2+\s2\t2\u2-16\s2\t2-
      5\s2\u2+115\s2+\nonumber\\
      &&\t2^3-18\t2^2-\t2\u2^2-3\t2\u2+
      166\t2+13\u2^2-65\u2-335)/\nonumber\\
        &&((\s2-1)^3(\u2-4))
\end{eqnarray}
\begin{eqnarray}
      K_{24}&=&8(2\s2^2\t2+2\s2^2\u2-24\s2^2+\s2\t2^2+\s2\t2
      \u2-22\s2\t2-11\s2\u2+\nonumber\\
     &&  217\s2+\t2^2+7\t2\u2+16
      \t2-2\u2^2+41\u2-511)/\nonumber\\
       &&((\s2-1)^2(\t2-1)(\u2-4))
\end{eqnarray}
\begin{eqnarray}
      K_{33}&=&-16(2\s2^2+2\s2\t2+6\s2\u2-46\s2+\t2^2\u2-10
      \t2^2-\nonumber\\
       &&4\t2\u2+74\t2+2\u2^3-22\u2^2+53\u2-118)/
 ((\s2-1)^2(\u2^2-4)^2)
\end{eqnarray}
\begin{eqnarray}
      K_{34}&=&-16(2(\s2^2+\s2\t2-11\s2+\t2^2+2\t2\u2-19\t2+\u2
      ^3-9\u2^2+28\u2+11))/\nonumber\\
       &&((\s2-1)(\t2-1)(\u2^2-4)^2)
\end{eqnarray}
\begin{eqnarray}
      K_{44}&=&-16(\s2^2\u2-10\s2^2+2\s2\t2-4\s2\u2+74\s2+2
      \t2^2+6\t2\u2-\nonumber\\
       &&46\t2+2\u2^3-22\u2^2+53\u2-118
      )/((\t2^-1)2(\u2^2-4)^2
\end{eqnarray}
\begin{eqnarray}
      K_{55}&=&4(\s2^3-\s2^2\t2+\s2^2\u2-14\s2^2-\s2\t2
     ^2+2\s2\t2\u2\nonumber\\
&& +20\s2\t2-\s2\u2^2 -8\s2\u2+40\s2+
      \t2^3+\t2^2\u2-14\t2^2\nonumber\\
&& -\t2\u2^2-8\t2\u2+40\t2+2
     \u2^3+4\u2^2-4\u2-168)/\nonumber\\
&&((\s2-1)^2(\t2-1)^2)
\end{eqnarray}
\begin{eqnarray}
      K_{51}&=&4(s2^3+\s2^2\t2-20\s2^2-\s2\t2^2+2\s2
      \t2\u2-12\s2\t2-\s2\u2^2\nonumber\\
&&       -4\s2\u2+168\s2-\t2^3-2 \t2^2\u2+24\t2^2+\t2\u2^2
          -12\t2\u2-88\t2+14\u2^2-232)\nonumber\\
&&      /((\s2-1)(\t2-1)^3)
\end{eqnarray}
\begin{eqnarray}
      K_{52}&=&-4(\s2^3+\s2^2\t2+2\s2^2\u2-24\s2^2-\s2
      \t2^2-2\s2\t2\u2+12\s2\t2\nonumber\\
&&       -\s2\u2^2+12\s2\u2+88
     \s2-\t2^3+20\t2^2+\t2\u2^2+4\t2\u2-168\t2-14\u2^2
      +232)\nonumber\\
&&      /((\s2-1)^3(\t2-1))
\end{eqnarray}
\begin{eqnarray}
      K_{53}&=&8(4\s2^2-\s2\t2^2+22\s2\t2+10\s2\u2-
      117\s2+\t2^3+2\t2^2\u2\nonumber\\
&&      -26\t2^2-\t2\u2^2+2\t2\u2
      +69\t2+4\u2^3-29\u2^2+34\u2+80)\nonumber\\
&&  /((\s2-1)^2(\t2-1)(\u2-4))
\end{eqnarray}
\begin{eqnarray}
       K_{54}&=&8(\s2^3-\s2^2\t2+2\s2^2\u2-26\s2^2+22
      \s2\t2-\s2\u2^2+2\s2\u2\nonumber\\
&&       +69\s2+4\t2^2+10\t2\u2-
      117\t2+4\u2^3-29\u2^2+34\u2+80)\nonumber\\
&&      /((\s2-1)(\t2-1)^2(\u2-4))
\end{eqnarray}
\newpage
\def\emline#1#2#3#4#5#6{%
       \put(#1,#2){\special{em:moveto}}%
       \put(#4,#5){\special{em:lineto}}}
\unitlength=1mm
\special{em:linewidth 1pt}
\linethickness{1pt}
\begin{center}
\begin{picture}(60.00,34.00)
\emline{10.00}{10.00}{1}{25.00}{10.00}{2}
\put(26.50,10.00){\oval(3.00,8.00)[]}
\emline{10.00}{30.00}{3}{12.00}{30.00}{4}
\emline{14.00}{30.00}{5}{17.00}{30.00}{6}
\emline{19.00}{30.00}{7}{22.00}{30.00}{8}
\put(26.50,29.50){\oval(3.00,9.00)[]}
\emline{25.00}{30.00}{9}{24.00}{30.00}{10}
\emline{27.00}{34.00}{11}{40.00}{34.00}{12}
\emline{28.00}{32.00}{13}{40.00}{32.00}{14}
\emline{28.00}{30.00}{15}{40.00}{30.00}{16}
\emline{27.00}{6.00}{17}{40.00}{6.00}{18}
\emline{28.00}{8.00}{19}{40.00}{8.00}{20}
\emline{28.00}{10.00}{21}{40.00}{10.00}{22}
\put(40.00,20.00){\circle*{5.20}}
\emline{28.00}{26.00}{23}{38.00}{22.00}{24}
\emline{28.00}{13.00}{25}{38.00}{18.00}{26}
\emline{42.00}{22.00}{27}{56.00}{26.00}{28}
\emline{42.00}{18.00}{29}{56.00}{14.00}{30}
\put(5.00,30.00){\makebox(0,0)[cc]{$\gamma$}}
\put(5.00,10.00){\makebox(0,0)[cc]{$p$}}
\put(60.00,26.00){\makebox(0,0)[cc]{$\J$}}
\put(60.00,14.00){\makebox(0,0)[cc]{$c$}}
\put(36.00,26.00){\makebox(0,0)[cc]{$c$}}
\put(36.00,13.00){\makebox(0,0)[cc]{$g$}}
\put(30.00,0.00){\makebox(0,0)[cc]{Fig.1}}
\end{picture}
\end{center}

\vspace{25mm}

\unitlength=1.00mm
\special{em:linewidth 1pt}
\linethickness{1pt}
\begin{picture}(126.00,107.37)
\emline{59.00}{69.09}{1}{69.00}{78.98}{2}
\emline{69.00}{78.98}{3}{69.00}{78.98}{4}
\emline{69.00}{78.98}{5}{84.00}{78.98}{6}
\put(85.00,83.93){\oval(2.00,9.89)[]}
\emline{85.00}{88.88}{7}{69.00}{88.88}{8}
\emline{69.00}{88.88}{9}{69.00}{98.77}{10}
\emline{69.00}{98.77}{11}{69.00}{98.77}{12}
\emline{69.00}{98.77}{13}{86.00}{98.77}{14}
\emline{69.00}{88.88}{15}{67.00}{86.73}{16}
\emline{67.00}{86.73}{17}{69.00}{86.73}{18}
\emline{69.00}{86.73}{19}{69.00}{86.73}{20}
\emline{69.00}{86.73}{21}{67.00}{85.01}{22}
\emline{67.00}{85.01}{23}{67.00}{85.01}{24}
\emline{67.00}{85.01}{25}{69.00}{85.01}{26}
\emline{69.00}{85.01}{27}{69.00}{85.01}{28}
\emline{69.00}{85.01}{29}{67.00}{82.85}{30}
\emline{67.00}{82.85}{31}{67.00}{82.85}{32}
\emline{67.00}{82.85}{33}{69.00}{82.85}{34}
\emline{69.00}{82.85}{35}{69.00}{82.85}{36}
\emline{69.00}{82.85}{37}{67.00}{81.13}{38}
\emline{67.00}{81.13}{39}{67.00}{81.13}{40}
\emline{67.00}{81.13}{41}{69.00}{81.13}{42}
\emline{69.00}{81.13}{43}{69.00}{81.13}{44}
\emline{69.00}{81.13}{45}{67.00}{78.98}{46}
\emline{67.00}{78.98}{47}{67.00}{78.98}{48}
\emline{67.00}{78.98}{49}{69.00}{78.98}{50}
\emline{86.00}{85.01}{51}{92.00}{85.01}{52}
\emline{92.00}{85.01}{53}{86.00}{85.01}{54}
\emline{86.00}{85.01}{55}{92.00}{85.01}{56}
\emline{86.00}{82.85}{57}{92.00}{82.85}{58}
\emline{91.00}{81.13}{59}{95.00}{83.71}{60}
\emline{95.00}{83.71}{61}{95.00}{83.71}{62}
\emline{95.00}{83.71}{63}{90.00}{86.73}{64}
\put(67.00,106.08){\makebox(0,0)[cc]{k}}
\put(67.00,69.09){\makebox(0,0)[cc]{q}}
\put(91.00,101.78){\makebox(0,0)[cc]{$q'$}}
\put(79.00,92.75){\makebox(0,0)[cc]{p}}
\put(79.00,73.82){\makebox(0,0)[cc]{p}}
\put(107.00,83.71){\makebox(0,0)[cc]{$p_J=2p$}}
\emline{94.00}{44.98}{65}{104.00}{44.98}{66}
\emline{94.00}{44.98}{67}{84.00}{35.09}{68}
\emline{104.00}{44.98}{69}{106.00}{42.83}{70}
\emline{106.00}{42.83}{71}{106.00}{44.98}{72}
\emline{106.00}{44.98}{73}{108.00}{42.83}{74}
\emline{108.00}{42.83}{75}{108.00}{44.98}{76}
\emline{108.00}{44.98}{77}{110.00}{42.83}{78}
\emline{110.00}{42.83}{79}{110.00}{44.98}{80}
\emline{110.00}{44.98}{81}{112.00}{42.83}{82}
\emline{112.00}{42.83}{83}{112.00}{44.98}{84}
\emline{112.00}{44.98}{85}{114.00}{42.83}{86}
\emline{114.00}{42.83}{87}{114.00}{44.98}{88}
\emline{104.00}{44.98}{89}{111.00}{53.15}{90}
\emline{114.00}{44.98}{91}{121.00}{53.15}{92}
\put(116.00,52.94){\oval(10.00,2.15)[]}
\emline{114.00}{44.98}{93}{123.00}{38.10}{94}
\emline{115.00}{54.01}{95}{120.00}{60.04}{96}
\emline{118.00}{54.01}{97}{122.00}{59.18}{98}
\emline{117.00}{59.18}{99}{123.00}{62.19}{100}
\emline{123.00}{57.88}{101}{123.00}{62.19}{102}
\emline{23.00}{41.53}{103}{51.00}{41.53}{104}
\emline{34.00}{41.53}{105}{32.00}{43.68}{106}
\emline{32.00}{43.68}{107}{34.00}{43.68}{108}
\emline{34.00}{43.68}{109}{32.00}{45.83}{110}
\emline{32.00}{45.83}{111}{34.00}{45.83}{112}
\emline{34.00}{45.83}{113}{32.00}{47.55}{114}
\emline{32.00}{47.55}{115}{34.00}{47.55}{116}
\emline{34.00}{47.55}{117}{32.00}{49.70}{118}
\emline{32.00}{49.70}{119}{34.00}{49.70}{120}
\emline{34.00}{49.70}{121}{32.00}{51.42}{122}
\emline{32.00}{51.42}{123}{51.00}{51.42}{124}
\put(51.00,46.48){\oval(2.00,9.89)[]}
\emline{52.00}{47.55}{125}{58.00}{47.55}{126}
\emline{52.00}{44.54}{127}{58.00}{44.54}{128}
\emline{56.00}{49.70}{129}{61.00}{45.83}{130}
\emline{56.00}{42.82}{131}{61.00}{46.69}{132}
\emline{32.00}{51.42}{133}{51.00}{62.61}{134}
\emline{94.00}{12.80}{135}{92.00}{14.95}{136}
\emline{92.00}{14.95}{137}{92.00}{14.95}{138}
\emline{92.00}{14.95}{139}{94.00}{14.95}{140}
\emline{94.00}{14.95}{141}{94.00}{14.95}{142}
\emline{94.00}{14.95}{143}{92.00}{17.10}{144}
\emline{92.00}{17.10}{145}{92.00}{17.10}{146}
\emline{92.00}{17.10}{147}{94.00}{17.10}{148}
\emline{94.00}{17.10}{149}{94.00}{17.10}{150}
\emline{94.00}{17.10}{151}{92.00}{18.82}{152}
\emline{92.00}{18.82}{153}{94.00}{18.82}{154}
\emline{94.00}{18.82}{155}{92.00}{20.97}{156}
\emline{92.00}{20.97}{157}{94.00}{20.97}{158}
\emline{94.00}{20.97}{159}{92.00}{22.69}{160}
\emline{92.00}{22.69}{161}{92.00}{22.69}{162}
\put(113.00,17.75){\oval(2.00,9.89)[]}
\emline{114.00}{18.82}{163}{121.00}{18.82}{164}
\emline{114.00}{15.81}{165}{121.00}{15.81}{166}
\emline{121.00}{15.81}{167}{121.00}{15.81}{168}
\emline{119.00}{14.09}{169}{126.00}{17.10}{170}
\emline{119.00}{20.97}{171}{126.00}{17.10}{172}
\emline{94.00}{12.80}{173}{113.00}{12.80}{174}
\emline{113.00}{22.69}{175}{83.00}{22.69}{176}
\emline{94.00}{12.80}{177}{107.00}{3.77}{178}
\emline{23.00}{16.80}{179}{51.00}{16.80}{180}
\emline{42.00}{27.12}{181}{51.00}{27.12}{182}
\put(52.00,21.96){\oval(2.00,10.32)[]}
\emline{53.00}{22.82}{183}{59.00}{22.82}{184}
\emline{53.00}{19.81}{185}{59.00}{19.81}{186}
\emline{56.00}{24.97}{187}{62.00}{21.10}{188}
\emline{62.00}{21.10}{189}{56.00}{18.09}{190}
\emline{33.00}{16.80}{191}{31.00}{18.95}{192}
\emline{31.00}{18.95}{193}{31.00}{18.95}{194}
\emline{31.00}{18.95}{195}{33.00}{18.95}{196}
\emline{33.00}{18.95}{197}{33.00}{18.95}{198}
\emline{33.00}{18.95}{199}{31.00}{21.10}{200}
\emline{31.00}{21.10}{201}{31.00}{21.10}{202}
\emline{31.00}{21.10}{203}{33.00}{21.10}{204}
\emline{33.00}{21.10}{205}{33.00}{21.10}{206}
\emline{33.00}{21.10}{207}{31.00}{22.82}{208}
\emline{31.00}{22.82}{209}{33.00}{22.82}{210}
\emline{33.00}{22.82}{211}{31.00}{24.97}{212}
\emline{31.00}{24.97}{213}{33.00}{24.97}{214}
\emline{33.00}{24.97}{215}{31.00}{26.69}{216}
\emline{31.00}{26.69}{217}{31.00}{26.69}{218}
\emline{32.00}{27.12}{219}{34.00}{24.97}{220}
\emline{34.00}{24.97}{221}{34.00}{27.12}{222}
\emline{34.00}{27.12}{223}{36.00}{24.97}{224}
\emline{36.00}{24.97}{225}{36.00}{27.12}{226}
\emline{36.00}{27.12}{227}{38.00}{24.97}{228}
\emline{38.00}{24.97}{229}{38.00}{27.12}{230}
\emline{38.00}{27.12}{231}{40.00}{24.97}{232}
\emline{40.00}{24.97}{233}{40.00}{27.12}{234}
\emline{40.00}{27.12}{235}{42.00}{24.97}{236}
\emline{42.00}{24.97}{237}{42.00}{27.12}{238}
\emline{22.00}{27.12}{239}{24.00}{24.97}{240}
\emline{24.00}{24.97}{241}{24.00}{27.12}{242}
\emline{24.00}{27.12}{243}{26.00}{24.97}{244}
\emline{26.00}{24.97}{245}{26.00}{27.12}{246}
\emline{26.00}{27.12}{247}{28.00}{24.97}{248}
\emline{28.00}{24.97}{249}{28.00}{27.12}{250}
\emline{28.00}{27.12}{251}{30.00}{24.97}{252}
\emline{30.00}{24.97}{253}{30.00}{27.12}{254}
\emline{94.00}{44.98}{255}{94.00}{47.13}{256}
\emline{94.00}{47.13}{257}{92.00}{44.98}{258}
\emline{92.00}{44.98}{259}{92.00}{47.99}{260}
\emline{92.00}{47.99}{261}{90.00}{45.84}{262}
\emline{90.00}{45.84}{263}{90.00}{47.99}{264}
\emline{90.00}{45.84}{265}{90.00}{48.85}{266}
\emline{90.00}{48.85}{267}{88.00}{47.13}{268}
\emline{88.00}{47.13}{269}{88.00}{51.00}{270}
\emline{88.00}{51.00}{271}{86.00}{48.85}{272}
\emline{86.00}{48.85}{273}{86.00}{51.86}{274}
\emline{86.00}{51.86}{275}{84.00}{50.14}{276}
\emline{84.00}{50.14}{277}{84.00}{53.15}{278}
\emline{103.00}{23.12}{279}{103.00}{25.27}{280}
\emline{103.00}{25.27}{281}{101.00}{23.12}{282}
\emline{101.00}{23.12}{283}{101.00}{26.13}{284}
\emline{101.00}{26.13}{285}{99.00}{23.98}{286}
\emline{99.00}{23.98}{287}{99.00}{26.13}{288}
\emline{99.00}{23.98}{289}{99.00}{26.99}{290}
\emline{99.00}{26.99}{291}{97.00}{25.27}{292}
\emline{97.00}{25.27}{293}{97.00}{29.14}{294}
\emline{97.00}{29.14}{295}{95.00}{26.99}{296}
\emline{95.00}{26.99}{297}{95.00}{30.00}{298}
\emline{95.00}{30.00}{299}{93.00}{28.28}{300}
\emline{93.00}{28.28}{301}{93.00}{31.29}{302}
\emline{40.00}{51.86}{303}{40.00}{54.01}{304}
\emline{40.00}{54.01}{305}{38.00}{51.86}{306}
\emline{38.00}{51.86}{307}{38.00}{54.87}{308}
\emline{38.00}{54.87}{309}{36.00}{52.72}{310}
\emline{36.00}{52.72}{311}{36.00}{54.87}{312}
\emline{36.00}{52.72}{313}{36.00}{55.73}{314}
\emline{36.00}{55.73}{315}{34.00}{54.01}{316}
\emline{34.00}{54.01}{317}{34.00}{57.88}{318}
\emline{34.00}{57.88}{319}{32.00}{55.73}{320}
\emline{32.00}{55.73}{321}{32.00}{58.74}{322}
\emline{32.00}{58.74}{323}{30.00}{57.02}{324}
\emline{30.00}{57.02}{325}{30.00}{60.03}{326}
\emline{42.00}{27.12}{327}{52.00}{34.86}{328}
\emline{69.00}{99.19}{329}{69.00}{101.34}{330}
\emline{69.00}{101.34}{331}{67.00}{99.19}{332}
\emline{67.00}{99.19}{333}{67.00}{102.20}{334}
\emline{67.00}{102.20}{335}{65.00}{100.05}{336}
\emline{65.00}{100.05}{337}{65.00}{102.20}{338}
\emline{65.00}{100.05}{339}{65.00}{103.06}{340}
\emline{65.00}{103.06}{341}{63.00}{101.34}{342}
\emline{63.00}{101.34}{343}{63.00}{105.22}{344}
\emline{63.00}{105.22}{345}{61.00}{103.06}{346}
\emline{61.00}{103.06}{347}{61.00}{106.08}{348}
\emline{61.00}{106.08}{349}{59.00}{104.35}{350}
\emline{59.00}{104.35}{351}{59.00}{107.37}{352}
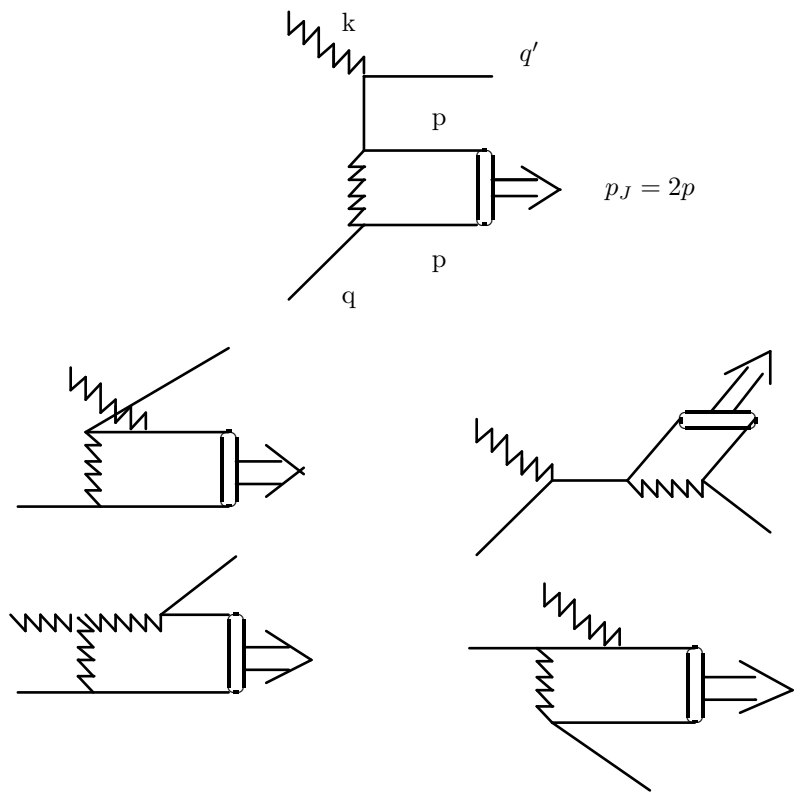
\centerline{Fig.~2}
\end{picture}

\vspace{5mm}

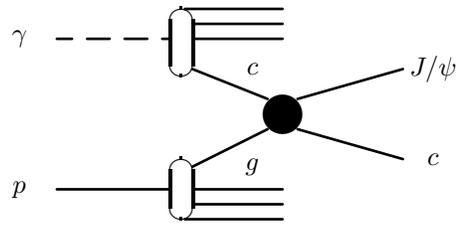
\end{document}